\documentclass[journal=jpcbfk,manuscript=article]{achemso}
\usepackage[version=3]{mhchem} 
\usepackage{graphicx}
\usepackage{amssymb}
\usepackage{epstopdf}

\newcommand{\onlinecite}[1]{\citenum{#1}\nocite{#1}}
\newcommand{\kT}{k_{\text{B}}T}

\newcommand{\dee}{\text{d}}

\renewcommand{\vec}[1]{\mathbf{#1}}
\newcommand{\vs}{\vec s}
\newcommand{\vr}{\vec r}

\newcommand{\vn}{\vec n}
\newcommand{\vx}{\vec x}
\newcommand{\del}{\partial}
\newcommand{\aTwiddle}{\tilde{a}}
\newcommand{\bTwiddle}{\tilde{b}}
\newcommand{\xTwiddle}{\tilde{x}}
\newcommand{\rhoTwiddle}{\tilde\rho}

\author{Patrick Varilly}
\altaffiliation{Current address: Department of Chemistry, University of
   Cambridge, Lensfield Road, Cambridge CB2 1EW, UNITED KINGDOM}
\author{David Chandler}
\email{chandler@cchem.berkeley.edu}
\affiliation[University of California, Berkeley]
{Department of Chemistry, University of California, Berkeley, California
  94720, U.S.A.}

\title{Water Evaporation: A Transition Path Sampling Study}

\begin{document}

\maketitle

\begin{abstract}
We use transition path sampling to study evaporation in the SPC/E model of
liquid water.  Based on thousands of evaporation trajectories, we
characterize the members of the transition state ensemble (TSE), which
exhibit a liquid-vapor interface with predominantly negative mean curvature
at the site of evaporation.  We also find that after evaporation is
complete, the distributions of translational and angular momenta of the
evaporated water are Maxwellian with a temperature equal to that of the
liquid.  To characterize the evaporation trajectories in their entirety, we
find that it suffices to project them onto just two coordinates: the
distance of the evaporating molecule to the instantaneous liquid-vapor
interface, and the velocity of the water along the average interface normal.
In this projected space, we find that the TSE is well-captured by a simple
model of ballistic escape from a deep potential well, with no additional
barrier to evaporation beyond the cohesive strength of the liquid.
Equivalently, they are consistent with a near-unity probability for a water
molecule impinging upon a liquid droplet to condense.  These results agree
with previous simulations and with some, but not all, recent experiments.

\end{abstract}

\newpage

\section{Introduction}

In a sample of water at equilibrium with its vapor, the rate of evaporation
is equal to the rate of condensation.  During condensation, not every gas
molecule that impinges on a liquid surface necessarily sticks.  The fraction
that does stick is known as the \emph{uptake coefficient}, $\gamma$, and by
microscopic reversibility, $\gamma$ can also be used to characterize
evaporation\cite{KolbEtAl:2010}.  Any deviation of $\gamma$ from~$1$ signals
some impediment to evaporation (or condensation) beyond the mere cohesive
strength of the liquid.  Measurements of~$\gamma$ have ranged from about
$0.001$~to~$1$ over the past century~\cite{EamesMarrSabir:1997}, but over
the last decade~\cite{DrisdellSaykallyCohen:2010}, they have been converging
to the range of $0.1$~to~$1$.  Li and coworkers\cite{LiEtAl:2001} measured
uptake of isotopically labeled water vapor in a train of water droplets to
obtain $\gamma = 0.17 \pm 0.03$ at $280\,$K, which increases with decreasing
temperature.  A similar result, $\gamma = 0.15 \pm 0.01$ at $282.5\,$K, was
obtained by Zientara and coworkers\cite{ZientaraEtAl:2008} from observations
of freely evaporating water droplets levitated in an electrodynamic trap.
Winkler and coworkers\cite{WinklerEtAl:2004, WinklerEtAl:2006}, on the other
hand, measured droplet growth in cloud chambers and claim to exclude values
of $\gamma < 0.4$ for temperatures below $290\,$K.  Their data is, in fact,
consistent with $\gamma = 1$ for temperatures ranging from
$250\,$K~to~$290\,$K.  Experiments done in the Saykally and Cohen
groups\cite{SmithEtAl:2006, DrisdellEtAl:2008, DrisdellSaykallyCohen:2009,
  DrisdellSaykallyCohen:2010}, which measure the drop in temperature as
water from a droplet in a droplet train evaporates into vacuum, indicate
that $\gamma = 0.62\pm0.09$ with little or no temperature dependence between
$245\,$K~and~$298\,$K.

The experimental uncertainty makes it unclear whether or not there is a
small barrier to evaporation.  To address this uncertainty, we were
motivated to carry out a detailed simulation study of evaporation using
transition path sampling~\cite{BolhuisEtAl:2002a} (TPS), a rare-event
sampling technique that can produce a statistically representative
collection of short evaporation trajectories with Boltzmann-distributed
(NVT) initial conditions and energy-conserving (NVE) dynamics.  Roughly
speaking, at $300\,$K, one water molecule evaporates from a $1\,$nm$^2$
patch of a liquid-vapor interface every $10\,$ns, which motivates using a
rare-event sampling technique. Other approaches could also be used to study
evaporation.  For example, in Ref.~\onlinecite{Mason:2011}, a single long
simulation of a small water droplet was performed at~$350\,$K, resulting in
$70$~evaporation events.  Another complementary approach is to study
condensation probabilities, since condensation is not rare at
all~\cite{TsurutaNagayama:2004, MoritaEtAl:2004,
  VieceliRoeselovaTobias:2004}.  A full discussion of the relationship
between evaporation and condensation trajectories is given in the Appendix.
The chief advantages of our approach are that we do not need to introduce
the approximation that the velocities and angular momenta of the evaporated
water molecule are Boltzmann-distributed, with a temperature equal to that
of the liquid, and that we are able to generate a large number of
evaporation trajectories (about~$5000$), which we can characterize
statistically instead of anecdotally.  Further, the framework for analyzing
TPS simulations can be used to obtain novel insight into evaporation
kinetics.

\section{Methods}

Throughout, we run simulations of liquid water with
LAMMPS~\cite{Plimpton:1995} using the SPC/E model of
water~\cite{BerendsenGrigeraStraatsma:1987}.  Lennard-Jones interactions are
truncated and shifted at a distance of~$10\,$\AA.  Electrostatic
interactions are calculated using the particle-particle particle-mesh (PPPM)
method~\cite{HockneyEastwood:1988}, with a relative error of $10^{-4}$.  The
bond and angle constraints of the water molecule are enforced using the
SETTLE algorithm~\cite{MiyamotoKollman:1992} to guarantee that trajectories
are time-reversible.  A timestep of $2\,$fs is used throughout.  In
simulations where we fix temperature, we use a Langevin thermostat with a
time constant of $2\,$ps.

We use the SPC/E model of water because it adequately captures a broad swath
of liquid water's properties.  With respect to bulk properties at $298\,$K,
its radial distribution function is quite accurate\cite{MarkNilsson:2001},
its density is within $1$\%\ of experiment\cite{VegaEtAl:2005}, its
compressibility\cite{MotakabbirBerkowitz:1990} of $4.1\times
10^{-10}\,$Pa$^{-1}$ is close to the experimental value of
$4.5\times10^{-10}\,$Pa$^{-1}$, and its dielectric
constant\cite{AragonesMacDowellVega:2011} of $70$ compares well with the
experimental value of~$78.2$.  The properties of its vapor-liquid transition
are also quite good: the model is explicitly parametrized to reproduce the
experimental enthalpy of vaporization\cite{BerendsenGrigeraStraatsma:1987},
its liquid-vapor surface tension is within about $10\,$\%\ of the
experimental value\cite{HuangGeisslerChandler:2001, VegadeMiguel:2007,
  PatelEtAl:2011a}, and its vapor pressure is within a factor of~$2$ of the
experimental value\cite{ErringtonPanagiotopoulos:1998}.  With regards to
transport properties, its self-diffusion coefficient\cite{MarkNilsson:2001}
of about $2.8\times 10^{-5}\,$cm$^2$/s compares well the experimentally
measured value of $2.3\times 10^{-5}\,$cm$^2$/s.  These properties lead us
to believe that the SPC/E model captures sufficient water-like behavior to
be useful in our study.  The model is not polarizable, but its
parametrization accounts implicitly for polarization in the bulk and results
in a semiquantitatively correct description of the liquid-vapor interface.
Of course, it is impossible to obtain arbitrarily precise quantitative
agreement with experiments using SPC/E or any other classical model of
water. However, the consistency of this model with general measures of
liquid-vapor coexistence, interfacial energetics, and molecular fluctuation
amplitudes and time scales gives us confidence in its qualitative
predictions about the molecular dynamics of water evaporation.

Transition path sampling~\cite{BolhuisEtAl:2002a} is used to generate nearly
$5000$ independent evaporation trajectories of length~$3\,$ps, which is long
enough to avoid spurious biases (see Supplementary Information).
Trajectories are constrained to start in a basin~$A$ in phase space,
corresponding to a condensed state, and end in a basin~$B$, corresponding to
an evaporated state.  In the analysis below, we consider only the
trajectories for which the system enters basin~$B$ after at least $2\,$ps to
avoid any biases towards unusually short evaporation trajectories.

Our explicit definitions of basins $A$~and~$B$ are as follows.  Basin~$A$
consists of all configurations where every water that is not hydrogen-bonded
to any other water is at most~$4\,$\AA\ away from the nearest
water (the ``position of a water'' means the position of its oxygen
  atom, unless otherwise stated).  Following
Ref.~\onlinecite{LuzarChandler:1996}, two waters are considered hydrogen bonded if
the distance between their oxygen atoms is below $3.5\,$\AA\ and the angle
between the OH~bond of the donor and the line connecting the two oxygen
atoms is below~$30^\circ$.  For our purposes, any other reasonable
definition of a hydrogen bond should yield nearly identical results.
Basin~$B$ is defined as all configurations of the system where there is
exactly one water molecule with no hydrogen-bonding partner that is more
than $8\,$\AA\ away from its nearest neighbor.  The distance cutoff used in
defining basin~$A$ is motivated by the extremely low likelihood for a water
in bulk to be that isolated.  We comment on our choice for the cutoff for
basin~$B$ below.

\begin{figure}
\setlength\tabcolsep{0pt}
\begin{center}
\includegraphics[height=10cm]{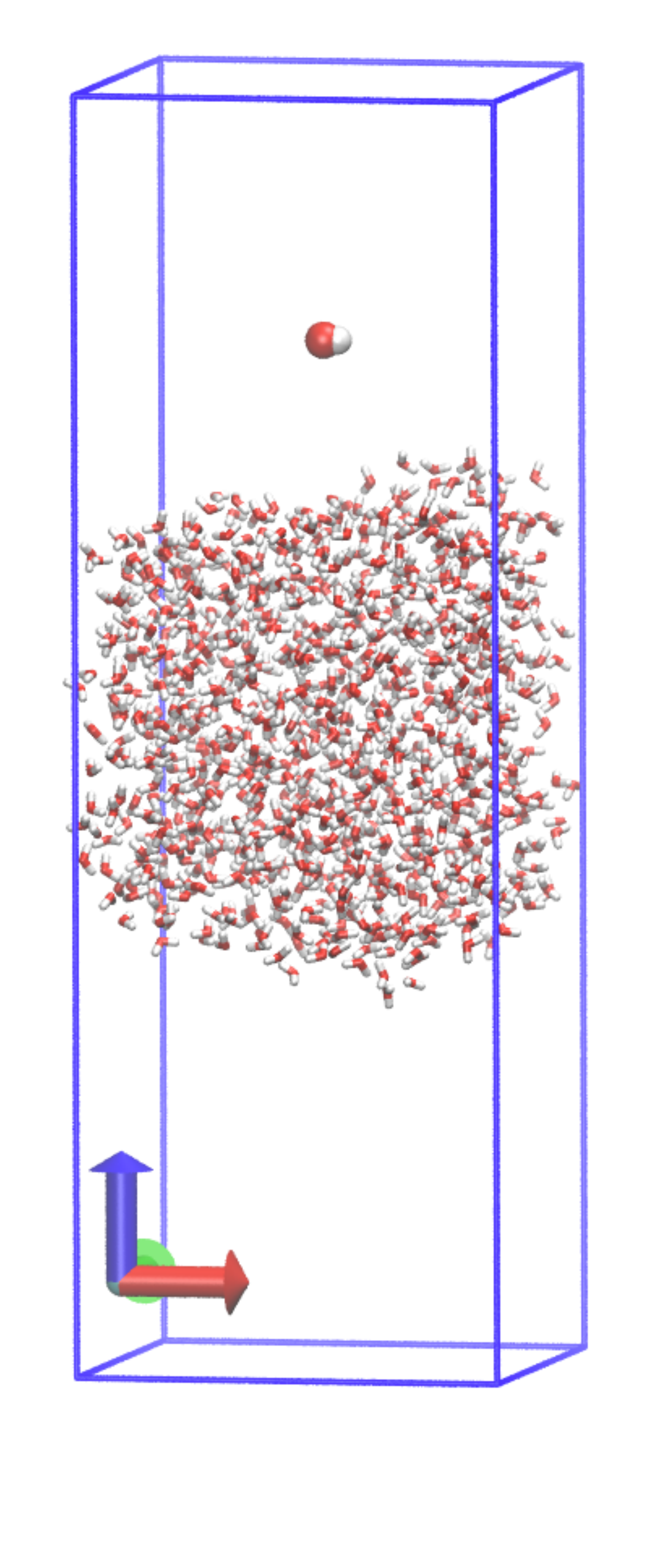}
\end{center}
\caption{\label{fig:setup}Snapshot of setup used to study water evaporation.}
\end{figure}

Transition path sampling is essentially a biased random walk in trajectory
space.  The initial trajectory of this walk is prepared as follows.  We
place water molecules in a crystalline arrangement in a
$30\times30\times30$\,\AA$^3$ periodic box so that the density of water
molecules matches the bulk density of water ($33.3$\,waters per nm$^3$,
equivalent to $0.997\,$g/ml), i.e., $900$~waters in total, and equilibrate
this system at $300\,$K for $50\,$ps.  Next, we enlarge the box to three
times its size along the $z$-dimension, and equilibrate the resulting system
for another~$50\,$ps.  At this stage, we have a
$30\times30\times90$\,\AA$^3$ periodic box containing a $30\,$\AA-thick slab
of water parallel to the $xy$-plane.  We then add a water molecule about
$15$\,\AA\ above the top of this slab with a random, thermal velocity.  An
example of the system at this stage is shown in Figure~\ref{fig:setup}.
Next, we evolve the system without a thermostat to yield a
$3$~to~$9\,$ps-long trajectory.  If in this time, the water molecule does
not condense (i.e., enter basin~$A$), we discard this initial trajectory and
start over.  Otherwise, we time-reverse the $3\,$ps stretch of the
trajectory immediately preceding condensation, and use this reversed
evaporation trajectory to seed the TPS random walk.  We have verified that
condensation fails to occur about $50\,$\%\ of the time, and that in all
cases is due to the water molecule having initial total momentum with a
positive $z$-component, so that the molecule moves away from the water slab
during the trajectory.  We have not observed any initial trajectory with a
water molecule initially headed towards the water slab and not condensing, a
fact that is consistent with a sticking coefficient~$\gamma$ of nearly~$1$,
as observed in previous similar simulations~\cite{TsurutaNagayama:2004,
  MoritaEtAl:2004, VieceliRoeselovaTobias:2004}.

The TPS random walk is performed as follows.  At every step, we choose to
make a shifting move $90\,$\% of the time, and a shooting move~$10\,\%$ of
the time, reflecting the low cost of shifting versus shooting.  In a
shifting move, we shift the trajectory forwards or backwards by a
time~$\Delta t$ uniformly distributed between $-1\,$~and~$1\,$ps.  Shooting
moves are performed as in the appendices of
Refs. \onlinecite{GeisslerDellagoChandler:1999}~and~\onlinecite{GeisslerChandler:2000}.
Briefly, the $3N$-dimensional vector of velocities weighted by the square
root of the atomic masses is rotated slightly, then projected down to a
hyperplane to enforce the constraints on velocities imposed by the fixed
bonds and angles of the water molecules.  The kinetic energy of the system
is then perturbed slightly. Generation and acceptance probabilities for this
move are chosen to satisfy detailed balance, and the magnitude of the
perturbations is chosen to yield an approximately $40\,$\%\ acceptance rate.

For each set of initial conditions, we performed between 10,000~and~20,000
TPS steps, recording a trajectory every 100 TPS steps.  Each recorded
trajectory is reasonably independent of the previous one, and the first $20$
recorded trajectories, which form the equilibration part of the random walk,
are discarded.  To further improve the sampling, we repeated the entire
procedure outlined in this section about $40$~times.  The final outcome of
this exercise is a set of $4696$ mostly uncorrelated evaporation
trajectories, with initial conditions drawn from a canonical ensemble at
temperature~$300\,$K and evolved in time with energy-conserving Newtonian
dynamics.

Our procedure induces a bias for evaporation trajectories where a single
water molecule comes off the liquid.  This bias arises from our definition
of basin~$B$ for the TPS random walk.  Before settling on this definition,
we explored the possibility of events where dimers or larger aggregates of
water evaporate as a unit, by using a more generous but cumbersome
definition of basin~$B$.  Specifically, a configuration was in basin~$B$ if
it contained exactly two separate clusters of waters, in each of which every
water was close to some other water in the cluster.  By observing the
evaporation events in these preliminary simulations, we convinced ourselves
that out of the rare events in which evaporation occurs, those involving
more than one water were far rarer still, so we neglected this possibility
in our final simulations in favor of using a simpler definition of
basin~$B$.

In analyzing the evaporation trajectories, it is useful to locate the
liquid-vapor interface at any instant in time, for which we use the method
of Ref.~\onlinecite{WillardChandler:2010}.  Briefly, we map a given
configuration of water oxygen atoms $\{\vr_i\}$ onto a smooth density field
$\rhoTwiddle(\vr)$ defined by the relation
\begin{equation}
\rhoTwiddle(\vr) = \sum_{i=1}^{N} \phi(|\vr - \vr_i|),
\end{equation}
where $N$ is the number of water molecules, and $\phi(r)$ is a Gaussian-like
smoothing function of width~$\xi=2.5\,$\AA\ (see Appendix).  The instantaneous
liquid-vapor interface is then defined implicitly as the set of
points~$\{\vs\}$ that satisfy
\begin{equation}
\rhoTwiddle(\vs) = (1/2) \rho_\ell,
\label{eqn:rhoTwiddle}
\end{equation}
where $\rho_\ell$ is the bulk density of liquid water.

\begin{figure}
\begin{center}
\includegraphics{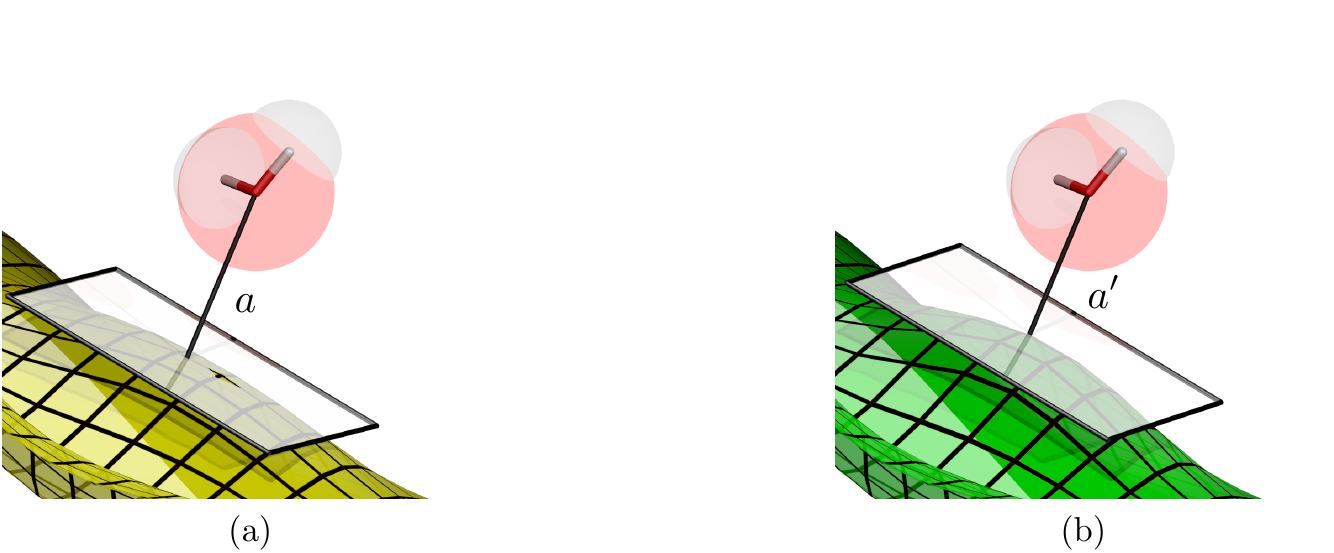}
\end{center}
\caption{\label{fig:Evap:EvapExample}Definition of water-to-surface
  distance~$a$ (or $a'$).  The plane shown is the tangent plane to the
  liquid-vapor interface at the point closest to the probe water molecule.
  In (a), the probe water is excluded from the definition of the interface;
  in (b), it is included.  These snapshots illustrate a typical interfacial
  deformation that accompanies an evaporation event.  The system is depicted
  at its transition state.}
\end{figure}

After locating the liquid-vapor interface, we follow
Ref.~\onlinecite{WillardChandler:2010} in defining a perpendicular
distance~$a$ from a probe water molecule at~$\vr$ to the interface, as
illustrated in Figure~\ref{fig:Evap:EvapExample}.  First, we locate the
point $\vs$ on the interface closest to~$\vr$, and calculate the
vapor-pointing normal vector to the interface there, $\hat{\vn}$.  Then $a$
is defined as the distance from $\vr$~to~$\vs$ projected along the
$\hat{\vn}$ direction,
\begin{equation}
a = \hat{\vn} \cdot (\vr - \vs).
\end{equation}
In Ref.~\onlinecite{WillardChandler:2010}, this distance was denoted
by~$a^*$.

An ambiguity arises about whether the probe molecule at~$\vr$ should or
should not be included when calculating the position of the liquid-vapor
interface.  Generally, we exclude it when calculating~$a$.  To discuss the
consequences of this choice, we define~$a'$ analogously to~$a$, but with the
probe water molecule included in the definition of the liquid-vapor
interface.

\section{Results}

\subsection{Evaporation correlates with negative mean curvature}
\label{sec:evap:curvature}

We first focus on the molecular details of the transition states of the
evaporation trajectories.  Ordinarily~\cite{BolhuisEtAl:2002a},
transition states are identified using committor functions.  The committor,
$p_B(\vx)$, of a spatial configuration~$\vx$ is defined as the fraction of
short trajectories that start at~$\vx$ with random thermal velocities, and
finish in basin~$B$.  At most points in a transition path, this function is
either $0$~or~$1$, with a quick crossover around the configurations that
dominate the dynamical bottleneck between $A$~and~$B$.  Thus, a pragmatic
definition of a transition state along a trajectory is the point where
$p_B(\vx) = 0.5$.

Implicit in the above definition of the committor function is the assumption
that momenta are not important in characterizing transition states.  In a
dense system, this assumption is generally true, since the velocity of any
particle decorrelates rapidly, usually within $1\,$ps~\cite{Chandler:1987}.
When examining evaporation, the assumption breaks down, since the velocity
of an evaporated water molecule decorrelates over much longer timescales.
The clearest manifestation of the problem is that the standard definition of
$p_B(t)$ leads to $p_B(t) \approx 0.5$ for a configuration containing a
single, clearly evaporated water molecule, since the water can likely
recondense if its rethermalized velocity points towards the liquid slab.

\begin{figure}
\begin{center}\includegraphics{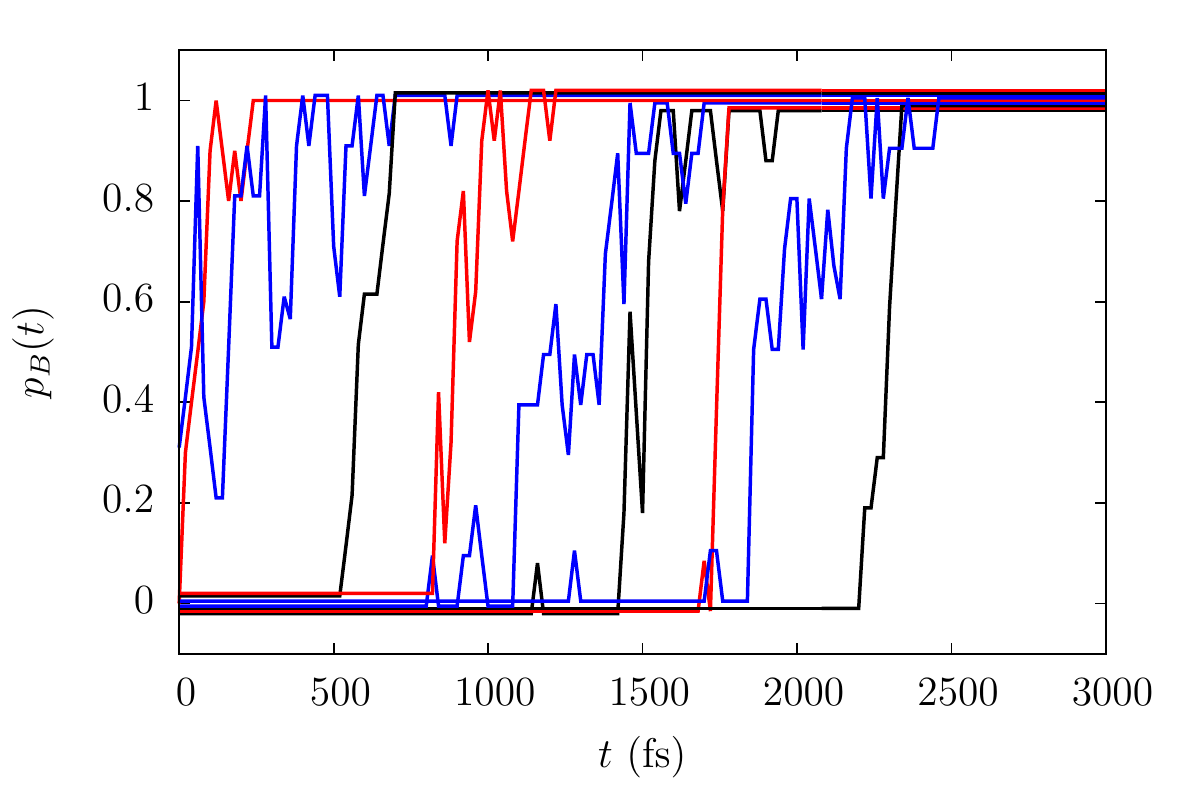}\end{center}
\caption{\label{fig:Evap:pB}Estimated committor, $p_B(t)$, sampled at
  $20\,$fs intervals for several evaporation trajectories.  The error in
  $p_B$ in the region around~$0.5$ is about~$0.15$.  To obtain this
  estimate, all velocities but the $z$-component of the evaporated water's
  velocity are randomized independently~$10$ times, after which a short
  trajectory is evolved forwards in time for up to $5\,$ps until the systems
  enters either basin~$A$ or basin~$B$.  For clarity, individual committors
  are slightly displaced vertically.}
\end{figure}

As a compromise, we have chosen to redefine the committor function to
include the $z$-component of the velocity of the evaporated water molecule.
Strictly speaking, it's impossible to tell which water molecule is ``the
evaporated molecule'' in an \emph{arbitrary} configuration, but this is not
a problem for identifying transition states along a transition path.
Figure~\ref{fig:Evap:pB} illustrates the typical behavior of $p_B(t) =
p_B(\vx(t),v_z^{\text{evap}}(t))$ defined in this way, estimated by spawning
$10$ short trajectories at every time point.

We have defined the transition states as the configuration at a time~$t_c$
equal to the mean of the first time for which $p_B(t)$ exceeds~$0.4$ and
the first time for which it exceeds~$0.6$.  The exact value of $t_c$ is not
very sensitive to the chosen cutoffs, as long as they are reasonable.  The
set of all configurations of the evaporation trajectories at their
respective times~$t_c$ comprises the transition state ensemble (TSE).

In many condensed-phase phenomena, collective coordinates are key.
Positions of individual atoms in the TSE do not by themselves appear
particularly remarkable or extraordinary.  Visual inspection confirms this
state of affairs in this particular case.  Instead, it is essential to
characterize the members of the TSE by looking for statistical trends in a
few collective coordinates.  Here, we focus on the instantaneous
liquid-vapor surface.  Let $\vs$ be the point on this surface that is
closest to the evaporating water molecule at any given time.  The mean
curvature, $H$, of the surface at~$\vs$ serves as a concise characterization
of collective fluctuations of water molecules at the liquid-vapor surface.
The mean curvature is defined as~\cite{Kreyszig:1991}
\begin{equation}
H = \frac{k_1 + k_2}{2},
\end{equation}
where $k_1$~and~$k_2$ are the principal curvatures at~$\vs$.  The magnitude
of a principal curvature is the reciprocals of the principal radius of
curvature, and its sign specifies whether the surface curves towards
(positive) or away (negative) from the normal direction along the
corresponding principal direction.  The mean curvature characterizes the
change in surface area upon infinitesimal deformation of the surface, so
it can be interpreted as a local characterization of the force of surface
tension on the liquid-vapor surface.  In particular, a deformation along the normal
direction by an infinitesimal distance $\epsilon$ changes the area element
$\dee A$ as~\cite{Kreyszig:1991}
\begin{equation}
\dee A \mapsto (1 - 2\epsilon H) \dee A.
\end{equation}

\begin{figure}
\begin{center}
\includegraphics{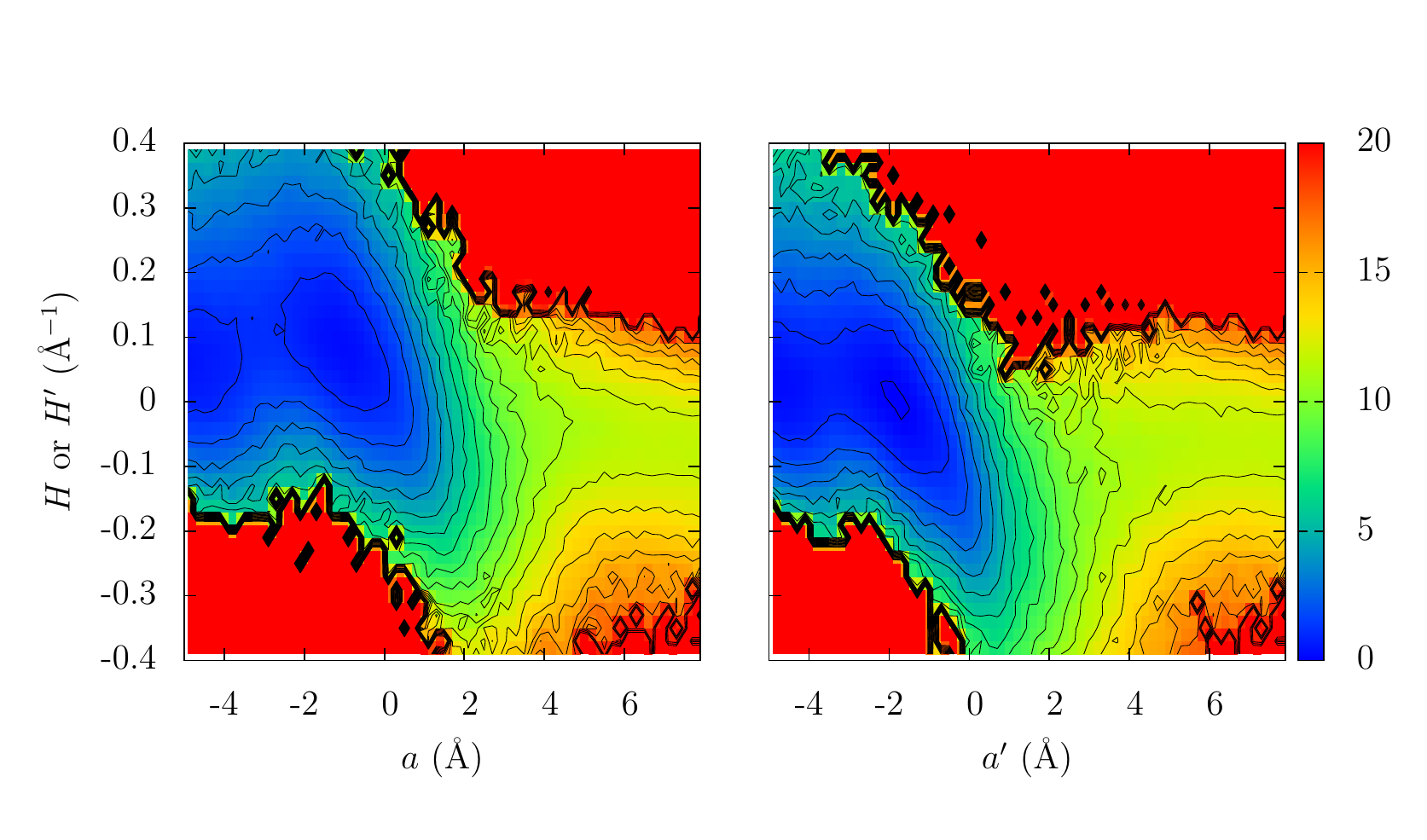}
\end{center}
\caption{\label{fig:Evap:KM}Free energy for height $a$ of a probe water
  molecule and the mean curvature, $H$, at the nearest point on the
  liquid-vapor surface (respectively $a'$~and~$H'$ if the probe water
  molecule is included in the definition of this surface).  Contours are
  spaced at~$1\,\kT$.}
\end{figure}

To establish a baseline, we first calculate the distribution of $H$ as a
function of the height~$a$ of a probe water molecule from the liquid-vapor
interface.  Figure~\ref{fig:Evap:KM} shows the results as a joint free
energy for $H$~and~$a$ (respectively, $H'$~and~$a'$ if the probe water
molecule is included in the definition of the liquid-vapor interface),
calculated using umbrella sampling as described in the Appendix.  At very
low and very high~$a$, only a trivial bias in $H$ is seen as a function
of~$a$, resulting from the nearest point on the surface being preferentially
one where the surface is bending most towards the probe water molecule.
However, an evident additional bias towards negative mean curvature can be
seen for $a$ just above the surface, indicating that a water molecule
suspended there significantly deforms the surface below it.
Figure~\ref{fig:Evap:EvapExample} shows an example of this kind of
deformation in one of the harvested evaporation trajectories.

\begin{figure}
\begin{center}
\includegraphics{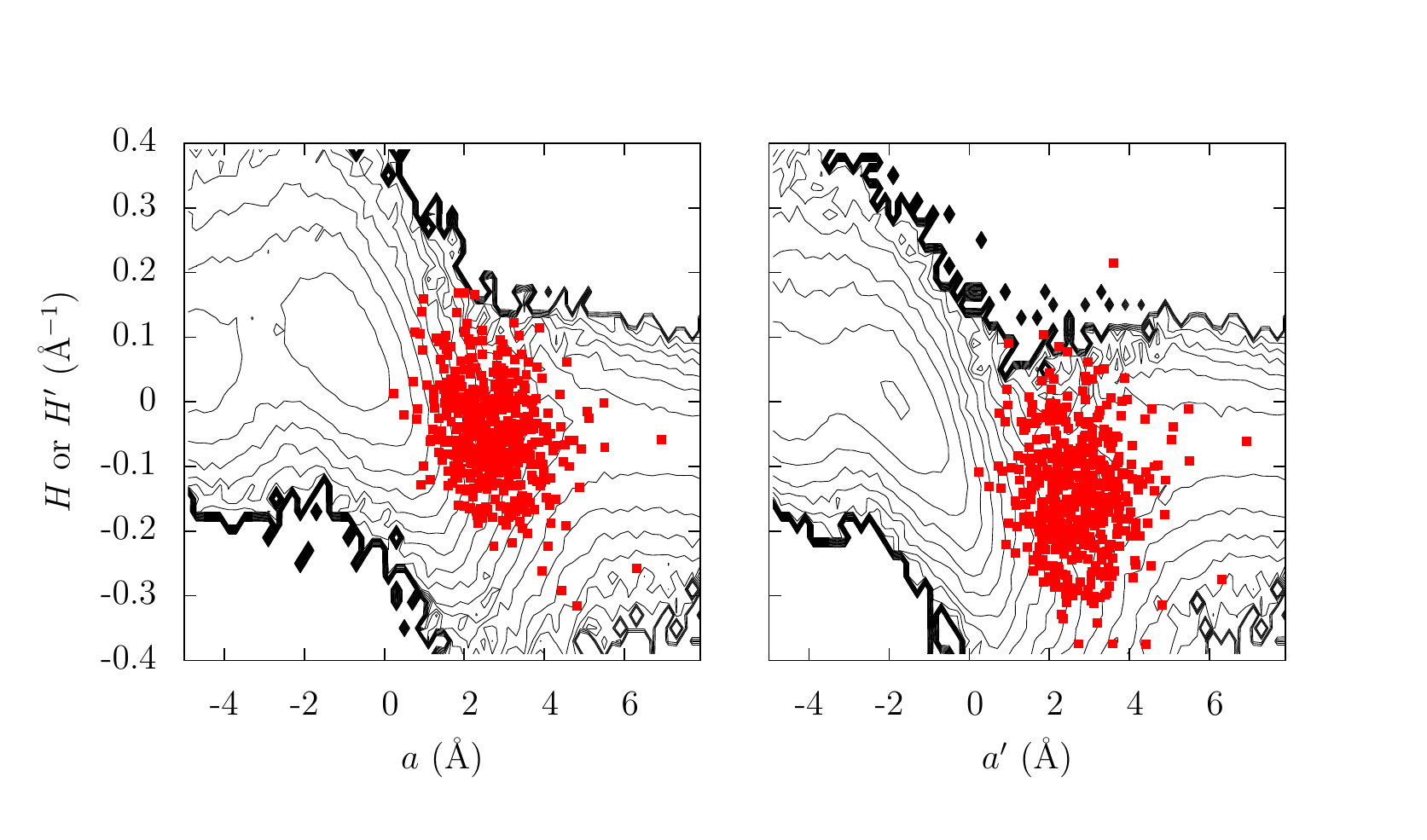}
\end{center}
\caption{\label{fig:Evap:KMTSE}Representative transition states of
  evaporation trajectories (red) projected onto the $H$~and~$a$
  coordinates.  The free energies of these coordinates are shown for
  comparison.  Labels as in Figure~\ref{fig:Evap:KM}.}
\end{figure}

Figure~\ref{fig:Evap:KMTSE} overlays the transition states of the
evaporation trajectories on the free energies of Figure~\ref{fig:Evap:KM}.
To a certain extent, the transition states exhibit some of the bias towards
negative curvature that can be seen in the equilibrium free energies.  The
bias is slight when the probe molecule is not included in the definition of
the liquid-vapor surface, but is clearer when the probe molecule is
included.  The definition of a liquid-vapor interface during the evaporation
process is somewhat ambiguous, and we regard full inclusion and full
exclusion as the two limiting extremes for a suitable definition.  Since the
bias towards negative curvature is present in both cases, our finding should
be robust with respect to reasonable changes in the definition of the
interface.

As discussed below, the preponderance of negative-mean-curvature
liquid-vapor interfaces does not correspond to an entropic barrier to
evaporation, but instead is a molecular manifestation of the cohesive
strength of the liquid.  Nevertheless, we anticipate that external
influences might be used to alter the microscopic details we describe, and
so may perhaps be used to exert control over evaporation.  Additionally, our
characterization establishes a baseline for understanding evaporation under
different conditions where barriers \emph{are} observed in simulations, such
as at higher temperatures~\cite{TsurutaNagayama:2004} or in the presence of
surfactants~\cite{TakahamaRussell:2011}.

\subsection{Post-evaporation momenta are Boltzmann-distributed}
\label{sec:Evap:momenta}

We now examine the center-of-mass velocities and angular momenta at the end
of each trajectory.  In all of the following results, we first estimate the
value of each observable independently in each TPS run, and then report the
mean of these values, with an error bar estimated as the standard error of
the mean.

\begin{figure}
\begin{center}
\includegraphics{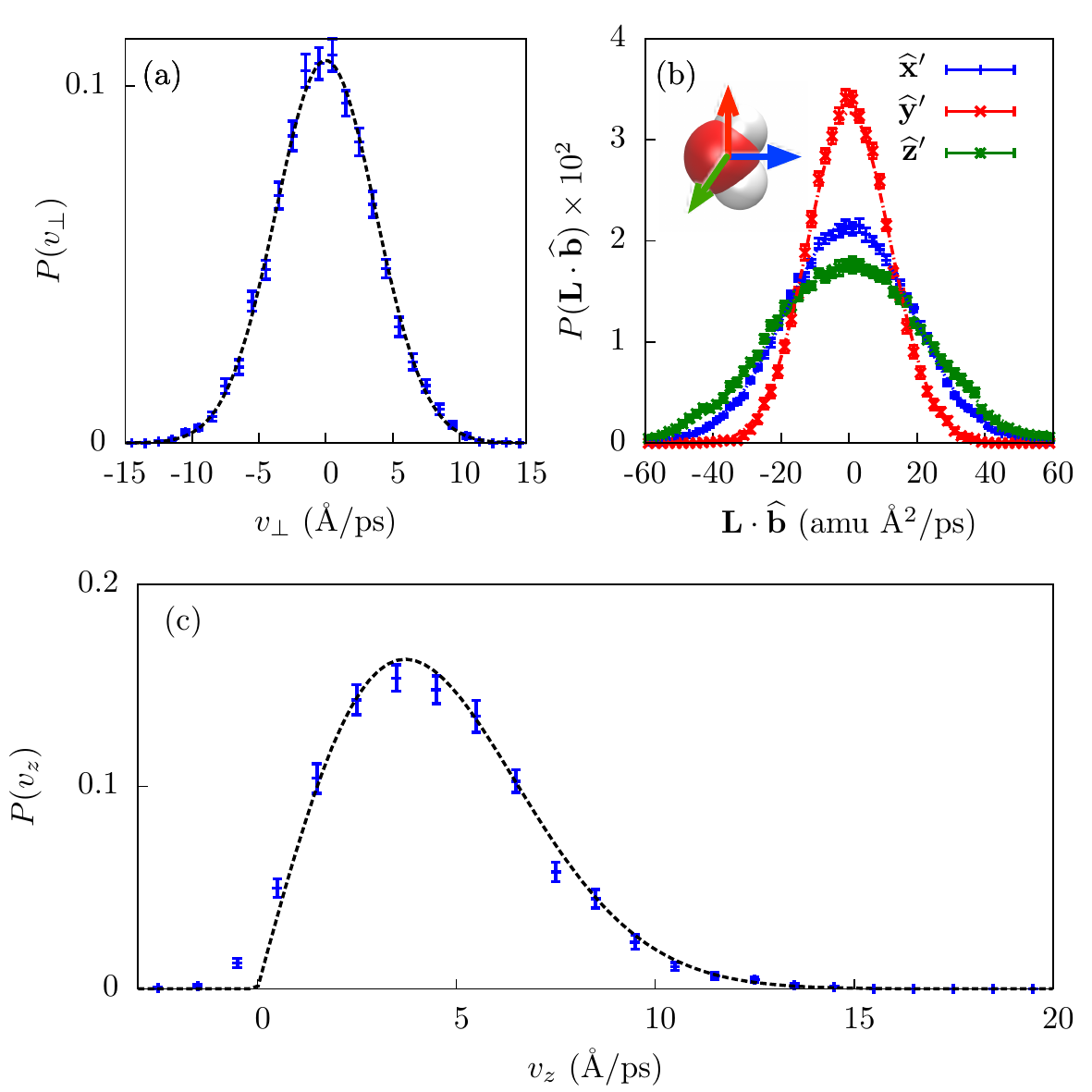}
\end{center}
\caption{\label{fig:Evap:Momenta}Distribution of (a) the component of
  center-of-mass velocity perpendicular to $\hat{\vec z}$; (b) the
  components of angular momentum of the evaporated water along the principal
  axes of inertia, measured at the end of an evaporation trajectory
  (symbols); and (c) the $z$-component of center-of-mass velocity.  For (a)
  and (b), the relevant Boltzmann distributions at temperature $T=300\,$K
  are also shown (dashed lines).  For (c), the expected result for thermal
  ideal gas particles evaporating from a deep, barrierless potential well is
  shown (Equation~\eqref{eqn:Evap:PostMomenta}, dashed line).}
\end{figure}

Figure~\ref{fig:Evap:Momenta}(a) shows the distributions of the
component of the evaporated water molecule's center-of-mass velocity along a
direction perpendicular to~$\hat{\vec z}$, measured at the end of an
evaporation trajectory.  Figure~\ref{fig:Evap:Momenta}(b) shows the
analogous distribution of the components of angular momenta along the
principal axes of inertia of the evaporating water molecule.  Both sets of
distributions are consistent with Boltzmann statistics at temperature~$T =
300\,$K.

The component of the velocity along the $z$~direction, $v_z$, has a more
interesting distribution, shown in Figure~\ref{fig:Evap:Momenta}(c).  We
enforce the constraint that water molecules first enter basin~$B$ with a
positive $v_z$ by flipping trajectories where this is not the case.  Hence,
no water molecules should have negative $v_z$ at the end of the evaporation
trajectory if the definition of basin~$B$ were sufficiently strict.  In
practice, the definition of basin~$B$ used here does not perfectly
discriminate between the evaporated states and states where recondensation
will occur.  Since the trajectories examined here are finite, a trajectory
where the system that transiently enters~$B$ before recondensing may appear
as an evaporation event, but with $v_z < 0$ at the end of the trajectory.
Only about 1\%\ of our trajectories exhibit this problem, which can in
principle be mitigated by using longer trajectories and a stricter
definition of basin~$B$.

\begin{figure}
\begin{center}\includegraphics{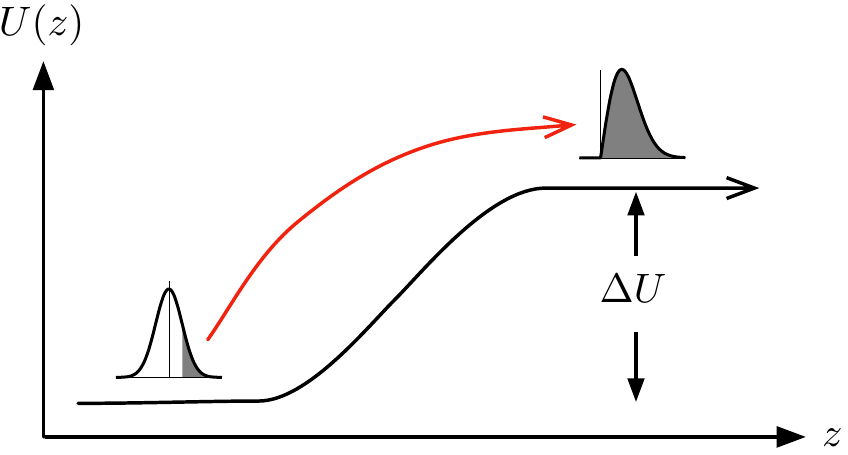}\end{center}
\caption{\label{fig:Evap:fromAWell}Ideal gas particles at the bottom of a
  deep, barrierless potential well have a Boltzmann distribution of
  velocities.  Only a fraction of particles have enough energy to escape the
  well.  After evaporating, but before thermalizing outside the well, the
  distribution of velocities of these particles is given by
  Equation~\eqref{eqn:Evap:PostMomenta}.}
\end{figure}

The expected distribution of $v_z$ for positive~$v_z$ can be deduced from a
simple model (Figure~\ref{fig:Evap:fromAWell}) of thermal ideal gas
particles evaporating from a deep, barrierless potential well of
depth~$\Delta U$.  Particles inside the well have a thermal distribution of
velocities, $P(v_i)$, given by
\begin{equation}
P(v_i) \propto \exp\left( -\frac12 \beta m v_i^2 \right).
\end{equation}
A particle with initial velocity velocity~$v_i$ can only escape the well if
$v_i$ is above a threshold velocity, $v_t$, given by $\frac12 m v_t^2 =
\Delta U$.  Were there a barrier, this threshold velocity would be higher,
but the remainder of this discussion would carry through unchanged. The
final velocity of this particle, $v_f$, is determined by conservation of
energy, independent of the details of any intermediate barrier:
\begin{equation}
\frac12 m v_i^2 = \frac12 m v_f^2 + \Delta U.
\label{eqn:Evap:Econs}
\end{equation}
This equation relates the distributions of initial and final velocities,
$P(v_i)$~and~$P(v_f)$ respectively, after correcting for the fact that for
finite trajectories, high initial velocities are overrepresented by a factor
of $|v_i|$, as there are proportionally more possible starting positions
compatible with the particle being outside the well at the end of the
trajectory.  The exact relationship is
\begin{equation}
P(v_f) \dee v_f \propto P(v_i) |v_i| \dee v_i,
\end{equation}
so
\begin{equation}
P(v_f) \propto P(v_i) |v_i| \frac{\dee v_i}{\dee v_f} \propto
\exp\left(-\frac12 \beta m v_i^2 \right) \frac{|v_i| \cdot v_f}{\sqrt{v_f^2 +
    \frac{2\Delta U}{m}}}.
\label{eqn:Evap:PostMomentaFull}
\end{equation}
Since the denominator in the last fraction is equal to $|v_i|$, we have
\begin{equation}
P(v_f) = \begin{cases}
\frac{m}{\kT} v_f \exp\left(-\frac12 \beta m v_f^2\right),&v_f > 0,\\
0,&v_f \leq 0.
\end{cases}
\label{eqn:Evap:PostMomenta}
\end{equation}
Were there a barrier of height $B$ to evaporation, the threshold $v_f$ above
would be $\sqrt{2 B / m}$ instead of~$0$, but the functional form would
remain unchanged.

While Equation~\eqref{eqn:Evap:PostMomenta} was derived for an ideal gas of
thermal particles escaping from a deep, barrierless potential well, it also
follows more generally from considerations of time reversibility (see
Appendix) and it describes the observed distribution of $v_z$ for
evaporating molecules surprisingly well (dashed line in
Figure~\ref{fig:Evap:Momenta}(c)).  In particular, low-velocity particles
are not underrepresented, which is consistent with there being no barrier to
evaporation.  A similar velocity distribution has been reported in
simulations of argon evaporation, which can be observed straightforwardly
without special sampling techniques like TPS~\cite{TsurutaNagayama:2004}.


\subsection{Potential of mean force for removing a water molecule from
  bulk is barrierless}
\label{sec:Evap:PMF}

\begin{figure}
\begin{center}\includegraphics{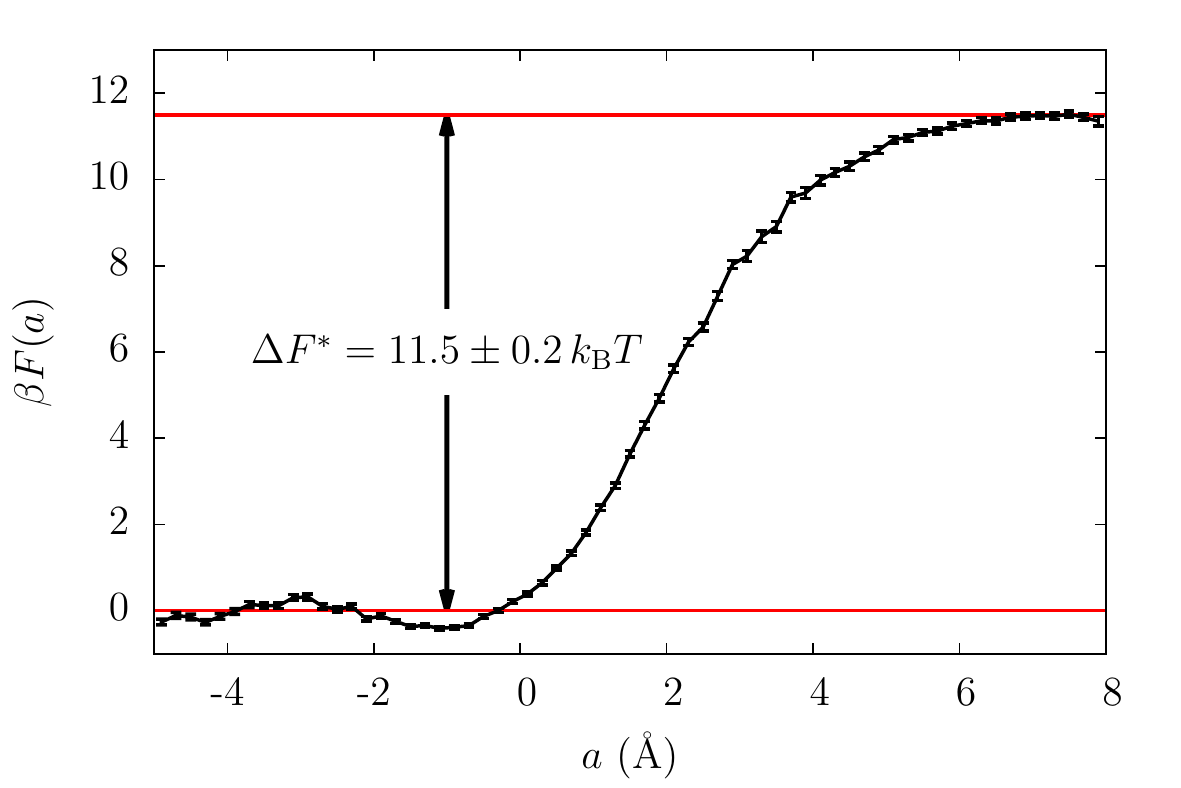}\end{center}
\caption{\label{fig:evap:pmf}Free energy for a single water molecule at a
  perpendicular height~$a$ from the liquid-vapor interface defined by the
  remaining water molecules.  The red lines are the free energies of the
  stable liquid and vapor phases, and are guides to the eye.  The biasing
  potentials used extend to $a = 7\,$\AA, so the apparent
  downturn at $a = 8\,$\AA\ is not statistically significant.}
\end{figure}

Figure~\ref{fig:evap:pmf} shows the free energy, $F(a)$, of an arbitrary
water molecule in our system as a function of the perpendicular distance to
the instantaneous liquid-vapor interface~$a$, calculated using umbrella
sampling (see Appendix).  Such a free energy profile is a reversible work or
a potential of mean force surface (i.e., its negative gradient is equal to
the mean force experienced by a water molecule along the~$a$
coordinate\cite{Chandler:1987}).  The essential feature of this free energy
is that it is barrierless.  Apart from density layering in the
bulk~\cite{WillardChandler:2010}, manifested as oscillations in $F(a)$ for
$a \lesssim 0\,$\AA, the bulk liquid simply sets up a deep potential well
for any individual water molecule, and a molecule in the vapor can simply
roll downhill into this well.  While the absence of a barrier along the $a$
coordinate does not preclude the existence of barriers along other
coordinates, we demonstrate below that the transition states of the
evaporation trajectories are consistent with $a$ describing the majority of
the evaporation reaction coordinate.

The depth of the well in $F(a)$, denoted by $\Delta F^*$, quantifies the
cohesiveness of the liquid with respect to the vapor.  Indeed, if we regard
a single water molecule as an independent particle moving in the potential
well~$F(a)$, then the relative density of this particle in the liquid,
$\rho_\ell$, with respect to that in the vapor, $\rho_g$, is given by
\begin{equation}
\rho_g = \rho_\ell e^{-\beta\Delta F^*}.
\end{equation}
We estimate from Figure~\ref{fig:evap:pmf} a value of $\Delta F^*$ of
$11.5\pm0.2\,\kT$.  This compares favorably with the value of $11.8\,\kT$
obtained by setting $\rho_g = P_{\text{vap}}/\kT$ and using the computed
value of $P_{\text{vap}}$ for SPC/E water at a temperature of~$300\,$K and
pressure of~$1\,$atm~\cite{ErringtonPanagiotopoulos:1998}.  For real water,
the analogous calculation yields $\Delta F^* = 10.5\,\kT$.

The range of $F(a)$ also characterizes the effective range of attraction
between a molecule in the vapor and the bulk slab, just under~$8\,$\AA.  It
is this range that motivates the definition of basin~$B$ described in the
Methods section.  Different models of water will have slightly different
ranges of attraction, but we do not expect discrepancies in the qualitative
behavior of $F(a)$.

Others have calculated a similar potential of mean force, but with respect
to the distance from the Gibbs dividing surface instead of the instantaneous
liquid surface, so that the details of the potential are masked by the
capillary wave fluctuations of the liquid-vapor interface.  Nevertheless,
their results for the SPC/E water model\cite{VachaEtAl:2004} and for a
polarizable water model due to Dang and Chang\cite{DangGarrett:2004} are
broadly similar to each other and to our own results.

\subsection{Transition states are consistent with diffusion out of a
  deep well}

\begin{figure}
\begin{center}\includegraphics{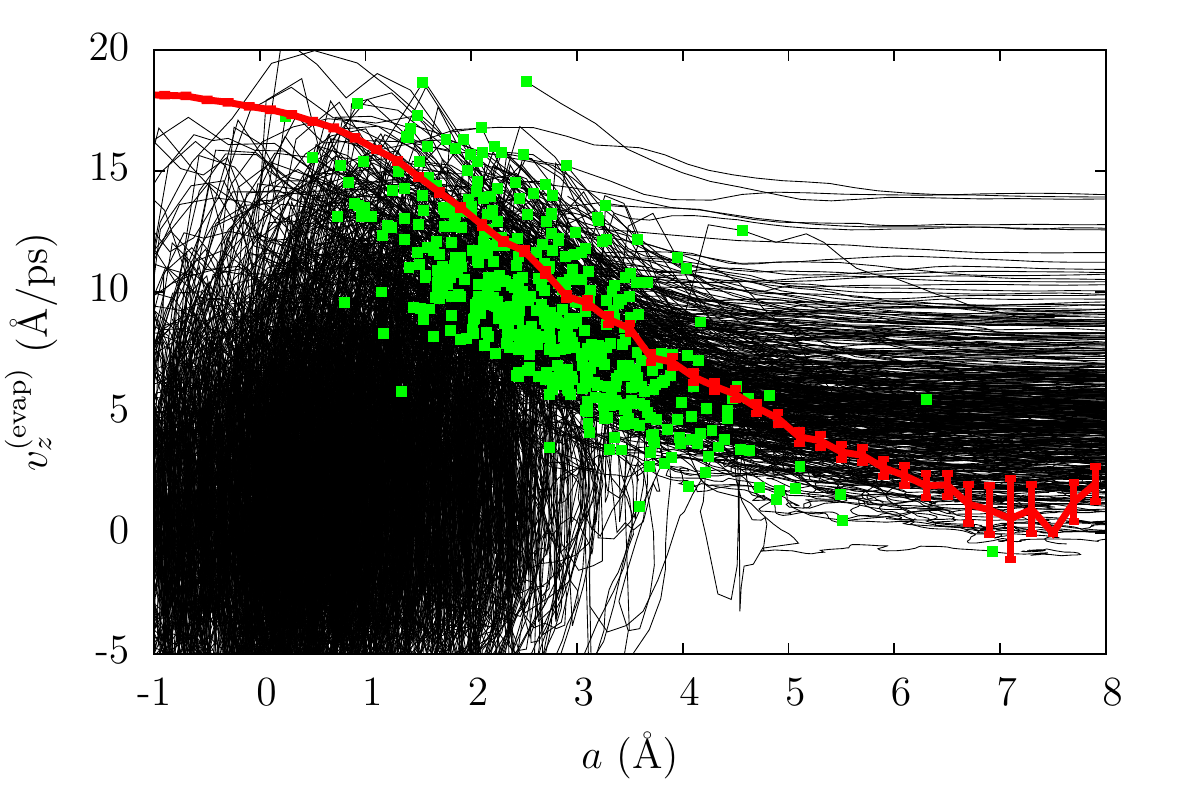}\end{center}
\caption{\label{fig:Evap:traceVzAStar}Evaporation trajectory traces
  projected onto variables $v_z^{\text{(evap)}}$~and~$a$ (black lines).
  The transition state of each trajectory, identified as described in the
  text, is highlighted by a green dot.  Red line: the expected transition
  state ensemble for a coarse model of ballistic escape from a potential
  shaped as in Figure~\ref{fig:evap:pmf}, given by
  Equation~\eqref{eqn:Evap:BallisticTSE}.}
\end{figure}

Figure~\ref{fig:Evap:traceVzAStar} depicts traces of many evaporation
trajectories projected onto the two coordinates
$v_z^{\text{(evap)}}$~and~$a$, with the transition state of each
trajectory highlighted in green.  Unlike similar traces onto many other
pairs of coordinates (not shown), there is a definite correlation between
the distance of the evaporated water from the liquid-vapor interface and its
speed in the $z$~direction.  We can partially rationalize this dependence by
conceiving of the free energy along~$a$ (Figure~\ref{fig:evap:pmf}) as an
actual potential energy well, and approximating the velocity along the $a$
direction with $v_z^{\text{(evap)}}$.  If evaporation were a ballistic
escape from this well, then the transition states would satisfy the
condition
\begin{equation}
\frac12 m (v_z^{\text{(evap)}})^2 = F(a).
\label{eqn:Evap:BallisticTSE}
\end{equation}
The points satisfying this relation are shown as a thick red line in
Figure~\ref{fig:Evap:traceVzAStar}.  Despite the evident crudeness of the
model, the transition states clearly cluster around the line of
Eq.~\eqref{eqn:Evap:BallisticTSE}.

\section{Discussion}

We have examined the process of evaporation of SPC/E water in detail, and
all the evidence suggests that there is no barrier to evaporation in this
model.  In other words, to evaporate, a water molecule near the surface only
needs to spontaneously acquire enough kinetic energy in the direction of the
liquid-vapor interface normal.  This view is consistent with the
distribution of $v_z$ for the final velocities (Figure~\ref{fig:Evap:Momenta}(c)),
the fact that the potential of mean force along a coordinate~$a$
perpendicular to the liquid-vapor surface is barrierless
(Figure~\ref{fig:evap:pmf}) and the fact that the transition states cluster
around values of $v_z$~and~$a$ that have a threshold amount of energy to
escape from the potential well set up by the remainder of the bulk
(Figure~\ref{fig:Evap:traceVzAStar}).  It is difficult to imagine
evaporation to be a mildly activated process and still be consistent with
these three pieces of evidence.

Our results are consistent with the near-unit condensation coefficient
measured in simulations in Refs.~\onlinecite{TsurutaNagayama:2004},
\onlinecite{MoritaEtAl:2004} and \onlinecite{VieceliRoeselovaTobias:2004},
but is in apparent contradiction with the most recent experimental
results\cite{SmithEtAl:2006, DrisdellEtAl:2008, DrisdellSaykallyCohen:2009,
  DrisdellSaykallyCohen:2010}, which suggest a barrier of around~$-\kT
\ln(\gamma) \approx 0.5\,\kT$.  The other experimental results cited in the
introduction suggest anything from the absence of a barrier to a barrier of
up to~$1.9\,\kT$.  Excluding the possibility that water molecules evaporate
as dimers, which would imply that an appreciable fraction of water molecules
in the vapor as dimerized (and recall that our preliminary transition path
sampling showed that there is not a significant fraction of SPC/E water
molecules that evaporate or condense as dimers or as larger clusters), such
large barriers should be clearly evident in direct simulations of water
condensation, but they are conspicuously absent\cite{TsurutaNagayama:2004,
  VieceliRoeselovaTobias:2004}.

The general lack of consensus between experiments~\cite{LiEtAl:2001,
  ZientaraEtAl:2008, WinklerEtAl:2004, WinklerEtAl:2006,
  DrisdellSaykallyCohen:2010} makes it unclear whether or not our result of
apparent unit evaporation coefficient agrees with reality, or if it is an
artifact of our simulations.  In particular, it could be that there is
indeed a barrier to evaporation and we cannot capture it, if that barrier
were due to fundamentally quantum effects.  By construction, these effects
are beyond the scope of the classical molecular dynamics simulations used
here.  Important quantum effects are plausible because librational motions
of water are strongly quantized: their typical wavenumbers,
around~$500\,$cm$^{-1}$, are comparable to the thermal wavenumber at $T =
300\,$K, around~$200\,$cm$^{-1}$.  More sophisticated simulation techniques
can incorporate many quantum effects at reasonable cost.  A notable
exception would be dynamical quantum coherence\cite{Miller:2012}, for which
a significant role would be surprising for intermolecular motions in a
strongly dissipitating system like liquid water.  If quantum effects were
limited to quantum dispersion and simple tunneling behavior, for instance,
one could explore the consequences of quantum uncertainty using ring-polymer
molecular dynamics~\cite{CraigManolopoulos:2004}.  However, our firm
expectation is that these more sophisticated simulations will produce
results that agree with those presented here, since generally, tunneling and
dispersion tend to lower effective barriers with respect to classical
expectations, not increase them.  Moreover, any account of such quantum
effects playing a dominant role would have to be compatible with the
observation\cite{DrisdellEtAl:2008} that the evaporation coefficient of
D$_2$O is equal, within errors, to that of H$_2$O.

Another possible source of discrepancy is our use of the SPC/E model of
water, and in particular, its lack of polarizability, which might result in
a qualitatively inaccurate description of events at the liquid-vapor
interface.  However, the agreement of its surface tension to the
experimental value (within about 10\%) suggests that the SPC/E model's
parameters implicitly capture enough detail about polarization to describe
the general mechanistic behavior of the liquid-vapor interface.  Moreover,
the addition of polarizability would likely reduce, not enhance, any
barriers to evaporation and/or condensation, since polarization induces an
additional attractive force between the liquid and a vapor molecule that is
relatively long-ranged.  Significantly, a previous study of direct
condensation that used the POL3 model of
water\cite{VieceliRoeselovaTobias:2004}, which is polarizable, is consistent
with $\gamma \approx 1$, i.e., barrierless evaporation.

Finally, extracting the evaporation coefficient from experiments involves
some interpretation and extrapolation, so it is conceivable that the quoted
results may be skewed by systematic errors that have not been accounted for.
For example, Morita et al.\cite{MoritaEtAl:2004} have previously argued that
Li et al's low reported evaporation coefficient\cite{LiEtAl:2001} may
actually be compatible with a value in the range of $0.2$~to~$1$ once the
effects of fluid flow on the gas surrounding their water droplet train are
corrected for.  As for the more recent experiments of
Refs.~\onlinecite{DrisdellSaykallyCohen:2010}, \onlinecite{SmithEtAl:2006},
\onlinecite{DrisdellEtAl:2008} and \onlinecite{DrisdellSaykallyCohen:2009},
these rely on a linear extrapolation of van't Hoff behavior of the Raman
spectrum of water down to supercooled temperatures in order to measure the
temperature of evaporating water droplets.  Recent Raman spectra of
magnetically trapped supercooled droplets, however, show that this
extrapolation may not be accurate~\cite{SuzukiEtAl:2012}.  This suggests
that the observed deviation from unit evaporation coefficient may also be in
part due to shortcomings in the calibration step of the experiments.  A
systematic error of 2\,\%\ in absolute temperature in the experiments (equal
to about 10\,\%\ in the temperature change during the course of the
measurements) would be sufficient to account for the discrepancy between the
experiments and our calculations.

\begin{acknowledgement}
The authors are grateful to Adam Willard, Amish Patel and Gil Nathanson for
helpful discussions, and to Walter Drisdell, Ron Cohen and Richard Saykally
for advice on many experiments devoted to this topic, including their own
work, which motivated this current paper.  In the early stages, P.V. was
supported by a Berkeley Fellowship, then by NIH Grant No.~R01-GM078102-04.
D.C. was also supported by the Director, Office of Science, Office of Basic
Energy Sciences, Materials Sciences and Engineering Division and Chemical
Sciences, Geosciences, and Biosciences Division of the U.S. Department of
Energy under Contract No. DE-AC02-05CH11231.  Both the raw evaporation
trajectories and the source code for the programs used in the analysis are
available upon request from the authors.
\end{acknowledgement}

\textbf{Supporting Information Available:} Distribution of times at which
trajectories leave basin~$A$ and enter basin~$B$, which support using
trajectories of length $3\,$ps.  This material is available free of charge
via the Internet at \verb+http://pubs.acs.org+.

\section{Appendix}

\subsection{Time reversibility and evaporation vs. condensation}

The observables measured using TPS for evaporation can be related to those
measured in simulations of condensation.  Let $h_A(x)$~and~$h_B(x)$ be
indicator functions of basins $A$~and~$B$.  They are equal to~$1$ if the
phase space point~$x$ is in the respective basin, and $0$~otherwise.  The
Boltzmann distribution, which specifies the initial conditions for our
evaporation trajectories, is denoted by $\rho(x)$.  The quantity $P_T[x\to
  y]$ is the probability density that a trajectory of length~$T$ has its
endpoint in the vicinity of~$y$, given that it started at $x$.  For the
energy-conserving dynamics that we use in the text,
\begin{equation}
P_T[x\to y] = \delta[y - U_T(x)],
\end{equation}
where $U_T(x)$ is the time evolution operator over a time~$T$.

The expectation of an observable $\mathcal{G}$ measured at the endpoint~$b$
of an evaporation trajectory of the kind sampled by TPS is given
by\cite{BolhuisEtAl:2002a}
\begin{equation}
\langle \mathcal{G}(b) \rangle_{\text{evap}}
= \frac{\int\dee b\, \int\dee a\, h_A(a) \rho(a) P_T[a\to b] h_B(b)
  \mathcal{G}(b)}
{\int\dee b\, \int\dee a\, h_A(a) \rho(a) P_T[a\to b] h_B(b)}.
\end{equation}
Conversely, the expectation of $\mathcal{G}$ measured at the beginning
point~$a$ of a condensation trajectory can be defined as follows:
\begin{equation}
\langle \mathcal{G}(b) \rangle_{\text{cond}}
= \frac{\int\dee b\, h_B(b) \rho(b) \mathcal{G}(b)}
{\int\dee b\, h_B(b) \rho(b)}.
\end{equation}
From these definitions, the following relation follows immediately:
\begin{equation}
\langle \mathcal{G}(b) \rangle_{\text{evap}}
= \frac{\left\langle \mathcal{G}(b) \int\dee a\, h_A(a) P_T[a\to
    b]\right\rangle_{\text{cond}}}
{\left\langle \int\dee a\, h_A(a) P_T[a\to
    b]\right\rangle_{\text{cond}}}.
\label{eqn:evapVsCond}
\end{equation}

Any configuration~$x$ can be mapped onto its time-reversed counterpart,
which we denote $\xTwiddle$, by inverting the direction of all the particle
momenta.  For time-reversible dynamics, such as that used in the text, we
have
\begin{equation}
P_T[a \to b] = P_T[\bTwiddle \to \aTwiddle],
\end{equation}
and further, for energy-conserving dynamics, if $a$~and~$b$ are in the same
trajectory, then
\begin{equation}
\rho(a) = \rho(b).
\end{equation}
With these relation, we can rewrite Eq.~\eqref{eqn:evapVsCond} in a more
usable form,
\begin{align}
\langle \mathcal{G}(b) \rangle_{\text{evap}}
&= \frac{\left\langle \mathcal{G}(b) \int\dee a\, P_T[\bTwiddle\to
    \aTwiddle] h_A(a)\right\rangle_{\text{cond}}}
{\left\langle \int\dee a\, P_T[\bTwiddle\to
    \aTwiddle] h_A(a)\right\rangle_{\text{cond}}},\\
&= \frac{\left\langle \mathcal{G}(\bTwiddle) \int\dee a\, P_T[b\to
    a] h_A(a)\right\rangle_{\text{cond}}}
{\left\langle \int\dee a\, P_T[b\to a] h_A(a)\right\rangle_{\text{cond}}}.
\label{eqn:evapVsCond2}
\end{align}
In the second equation, we have renamed the integration variables
$a$~and~$b$, and exploited that $h_A(a) = h_A(\aTwiddle)$ and $h_B(b) =
h_B(\bTwiddle)$.

Equation~\eqref{eqn:evapVsCond2} tells us that averages over TPS
trajectories are equivalent to time-reversed averages over trajectories that
start in B and end in basin~$A$ after time~$T$.  A priori, there is no
requirement that the water that is condensing have an initial velocity that
is directed towards the liquid slab, though trajectories that do not satisfy
this condition are very unlikely to end in basin~$A$.

A subtle point about Equation~\eqref{eqn:evapVsCond2} is that the
conditional factor $\int\dee a\, P_T[b\to a] h_A(a)$ cannot be approximated
as~$1/2$ for large~$T$.  Indeed, basin~$B$ is potentially unbounded, so no
matter how large a $T$ is chosen, there will be configurations in~$B$ with
an initial velocity of the isolated water is too low for the system to
escape basin~$B$ in time~$T$.  There are two potential solutions to this
problem.  One is to make basin~$B$ finite.  Alternately, and more
revealingly, one can model the consequence of the unboundedness of
basin~$B$, as we do below.

For concreteness, we consider a simpler definition of $B$ than the one used
in the text, which is easier to analyze and allows us to make the connection
between the discussion here and kinetic rate theory\cite{Chandler:1978}.
Let $z(b)$ be the $z$-coordinate of the evaporated water molecule's center
of mass, and let $v_z(b)$ be the corresponding component of the velocity.
The simpler basin~$B$ consists of all configurations~$b$ for which $z(b) >
z^*$.  With this definition, we can make the following approximation:
\begin{equation}
P_T[b\to a] h_A(a) \approx \Theta[|v_z(b)| T - (z(b) - z^*)] P_\tau[b^* \to a],
\end{equation}
with $\tau \ll T$ a small, fixed time and $b^*$ the point along the
trajectory starting at $b$ where $z$ is first equal to $z^*$.  In other
words, the probability for a configuration $b$ to end in basin~$A$ is mostly
determined by whether $T$ is long enough to get to the boundary of~$B$, and
then a kinetic factor that's virtually independent of $T$.  We also assume
that $\mathcal{G}(b)$ is independent of $z(b)$, so we can replace
$\mathcal{G}(b)$ by $\mathcal{G}(b^*)$.  Since the mapping from $b$~to~$b^*$
is area preserving, we have
\begin{equation}
\langle \mathcal{G}(b) \rangle_{\text{evap}} \approx \frac{\left\langle
  \mathcal{G}(\bTwiddle) |v_z(b)| \delta[z(b) - z^*] \int\dee a\,
  P_\tau[b\to a] h_A(a)\right\rangle_{\text{cond}}} {\left\langle |v_z(b)|
  \delta[z(b) - z^*] \int\dee a\, P_\tau[b\to a]
  h_A(a)\right\rangle_{\text{cond}}}.
\label{eqn:evapVsCondLongT}
\end{equation}

In comparison, the transmission coefficient for a reaction from $B$~to~$A$
after a transient time~$\tau$ is given by\cite{Chandler:1978}
\begin{equation}
\kappa_{B\to A}(\tau) = \frac{\left\langle |v_z(b)| \delta[z(b) - z^*]
  \int\dee a\, P_\tau[b\to a] h_A(a)\right\rangle_{\text{cond}}}
      {\left\langle |v_z(b)| \delta[z(b) - z^*] \cdot (1/2)\right\rangle_{\text{cond}}}.
\end{equation}
As is normal in reaction rate calculations, this transmission coefficient is
almost independent of $\tau$ for values of $\tau$ greater than molecular
timescales but smaller than implied by typical reaction rates.  Here, those
conditions require that $1\,\text{ps} \ll \tau \ll 1\,\text{ns}$.  In this
plateau regime, the transmission coefficient is equal to the uptake
coefficient, $\gamma$.  If this coefficient is~$1$ and $z(b)$ is high
enough that an initially evaporating water molecule does not recondense,
then we have
\begin{equation}
\int\dee a\, P_\tau[b\to a] h_A(a) \approx \Theta[-v_z(b)],
\end{equation}
so that
\begin{equation}
\langle \mathcal{G}(b) \rangle_{\text{evap}} \approx \frac{\left\langle
  \mathcal{G}(\bTwiddle) |v_z(b)| \delta[z(b) - z^*]
  \Theta[-v_z(b)]\right\rangle_{\text{cond}}}
{\left\langle |v_z(b)|
  \delta[z(b) - z^*] \Theta[-v_z(b)]\right\rangle_{\text{cond}}}.
\label{eqn:evapVsCondLongTAlpha1}
\end{equation}

The quantity on the right-hand sides of Equations
\eqref{eqn:evapVsCondLongT}~and~\eqref{eqn:evapVsCondLongTAlpha1} is what is
directly measured in condensation simulations.  Obtaining them required
several assumptions, all of which are reasonable in the context of this
paper.  However, our treatment here highlights the assumptions explicitly,
and will be useful in contexts where these assumptions may not apply.

One simple application of Equations
\eqref{eqn:evapVsCondLongT}~and~\eqref{eqn:evapVsCondLongTAlpha1} is to
calculate the distribution of $v_z$ for the evaporated water molecules.
Substituting $\mathcal{G}(b) = \delta[v_z(b) - v_f]$ immediately yields
Equation~\eqref{eqn:Evap:PostMomentaFull}.

\subsection{Choice of density smoothing function $\phi(r)$}

In the main text, the liquid-vapor interface is defined as an isosurface of
the smoothed density field~$\rhoTwiddle(\vr)$, constructed by convoluting
the instantaneous water density (a sum of Dirac delta functions) with a
smoothing kernel, $\phi(r)$.  In Ref.~\onlinecite{WillardChandler:2010}, the
choice for $\phi(r)$ was a Gaussian of width~$\xi$, truncated and shifted
at~$r = 3\xi$.  Since our study focuses on the curvature of the liquid-vapor
interface, the discontinuity in first and second derivatives of $\phi(r)$ at
the cutoff point is inconvenient.  Instead, to ensure that
$\rhoTwiddle(\vr)$ is sufficiently smooth, we use a $\phi(r)$ that results
from stitching two cubic functions of $r$ at the point $r = c$, subject to
the following conditions: (a) $\phi(r)$, $\phi'(r)$~and~$\phi''(r)$ are
continuous at $r = c$, (b) $\phi(3\xi) = 0$, (c) $\phi'(0) = \phi'(3\xi) =
0$, (d) $\phi''(3\xi) = 0$, (e) $\int_0^\infty \dee r\, 4\pi r^2 \phi(r) =
1$.  These eight conditions uniquely specify $\phi(r)$.
The stitching point is chosen empirically to be $c = 2.1\xi$ so that
$\phi(r)$ closely resembles a Gaussian with standard deviation~$\xi$.  In
Ref.~\onlinecite{WillardChandler:2010}, a value of $\xi=2.4\,$\AA\ was
chosen, which leads to about~$7\,$\%\ of our trajectories having an
ambiguous liquid-vapor interface at some timestep (i.e.,
Eq.~\eqref{eqn:rhoTwiddle} defining more than two liquid-vapor interfaces).  We
have found it convenient to use a slightly higher value, $\xi=2.5\,$\AA,
whereby the fraction of trajectories with ambiguous liquid-vapor interfaces
at any timestep drops to about~$3\,$\%.  For simplicity, all of these
trajectories are discarded in their entirety in the analyses above.

\subsection{Umbrella sampling with respect to the position of the liquid-vapor interface}

We have used umbrella sampling to collect statistics on rare configurations
of our system where a probe water molecule is at a fixed perpendicular
distance~$a$ (or~$a'$) from the instantaneous liquid-vapor interface.  To do
this, we have used the indirect umbrella sampling method (INDUS) that we
have previously used in different contexts\cite{PatelEtAl:2011b}.  Briefly,
we umbrella sample along a different coordinate that tracks $a$, use
MBAR\cite{ShirtsChodera:2008} to properly reweight all our samples, then
compute histograms for $a$ and possibly other variables from these weighted
samples.  The coordinate we use is the distance~$\aTwiddle$ from the probe
water molecule to the instantaneous liquid-vapor interface directly below
it.  Let $h(x,y; \vr^N)$ be the $z$-coordinate of the liquid-vapor interface
with the given values of $x$~and~$y$, which in turn depends on the positions
of the $N$~water oxygen atoms.  The umbrella potential we use is
\begin{equation}
V(\vr^N) = \frac{\kappa}{2} \bigl[ z_n - h(x_n, y_n; \vr^N) - \aTwiddle\bigr]^2.
\label{eqn:umbrellaWillard}
\end{equation}
Here, $n$ is the index of the probe water molecule, with coordinates $(x_n,
y_n, z_n)$.  The value of~$h(x_n,y_n; \vr^N)$ is defined implicitly by the
equation
\begin{equation}
\rhoTwiddle\bigl( x_n, y_n, h(x_n,y_n; \vr^N); \vr^N \bigr) = (1/2)\rho_\ell.
\label{eqn:nonboltz:WillardUmbrellaH}
\end{equation}
We henceforth suppress the dependence of it on~$\vr^N$.  In a slab of water,
there are usually two disjoint interfaces at the slab's top and bottom, so
this equation has two solutions.  For concreteness, we always refer to the
top interface of the slab.

To calculate~$h(x_n, y_n)$ quickly at every timestep, as well as its
gradient with respect to particle positions, we note that the value of
$h(x_n, y_n)$ at one timestep is similar to its value at the next timestep.
We have thus implemented a parallel Newton-Raphson solver to calculate
$h(x_n, y_n)$, with the starting guess at one timestep equal to the value
of~$h(x_n, y_n)$ at the previous timestep.  In a typical simulation,
convergence to $10^{-3}\,$\AA\ occurs after just one or two iterations.

To calculate the forces implied by the umbrella potential, we need to
calculate the gradient of Equation~\eqref{eqn:umbrellaWillard} with respect
to particle positions.  We present explicit expression below, where $h$~and
its derivatives are evaluated at~$(x_n,y_n)$, while $\rhoTwiddle$~and its
derivatives are evaluated at~$(x_n,y_n,h(x_n,y_n))$.  To simplify the
calculation, we assume that the tagged particle~$n$ is not a water oxygen,
and then relax this restriction.  By taking the total derivative of
Equation~\eqref{eqn:nonboltz:WillardUmbrellaH} with respect to the position
of oxygen atom~$i$, we find that
\begin{equation}
\frac{\dee(\rhoTwiddle - \rho_\ell/2)}{\dee\vr_i} = \frac{\del\rhoTwiddle}{\del z}
\frac{\dee h}{\dee\vr_i} + \frac{\del\rhoTwiddle}{\del \vr_i} = 0.
\end{equation}
Hence,
\begin{equation}
\frac{\dee h}{\dee\vr_i} = - \frac{\del\rhoTwiddle}{\del \vr_i} \bigg/
\frac{\del\rhoTwiddle}{\del z}.
\label{eqn:nonboltz:gradih}
\end{equation}
The derivative with respect to
the position of particle~$n$ is obtained similarly, so
\begin{subequations}
\begin{align}
\frac{\dee(\rhoTwiddle - \rho_\ell/2)}{\dee{x_n}} =
\frac{\del\rhoTwiddle}{\del x} + \frac{\del\rhoTwiddle}{\del z}
\frac{\dee h}{\dee{x_n}} &= 0,\\
\frac{\dee(\rhoTwiddle - \rho_\ell/2)}{\dee{y_n}} =
\frac{\del\rhoTwiddle}{\del y} + \frac{\del\rhoTwiddle}{\del z}
\frac{\dee h}{\dee{y_n}} &= 0,\\
\frac{\dee(\rhoTwiddle - \rho_\ell/2)}{\dee{z_n}} &= 0.
\end{align}
\end{subequations}
 Hence,
\begin{subequations}
\begin{align}
\frac{\dee h}{\dee{x_n}}
&= - \frac{\del\rhoTwiddle}{\del x} \bigg/ \frac{\del\rhoTwiddle}{\del z},\\
\frac{\dee h}{\dee{y_n}}
&= - \frac{\del\rhoTwiddle}{\del y} \bigg/ \frac{\del\rhoTwiddle}{\del z},\\
\frac{\dee h}{\dee{z_n}} &= 0.
\end{align}
\label{eqn:nonboltz:gradnh}
\end{subequations}
If the probe water molecule~$n$ is itself included in the definition of the
liquid-vapor interface, then $\dee h/\dee \vr_n$ is the sum of the
right-hand sides of Equations
\eqref{eqn:nonboltz:gradih}~and~\eqref{eqn:nonboltz:gradnh}.

\section{Supplementary Information: Length of evaporation trajectories}

In this section, we show that the $3\,$ps length of our TPS trajectories is
long enough.

\begin{figure}
\begin{center}\includegraphics{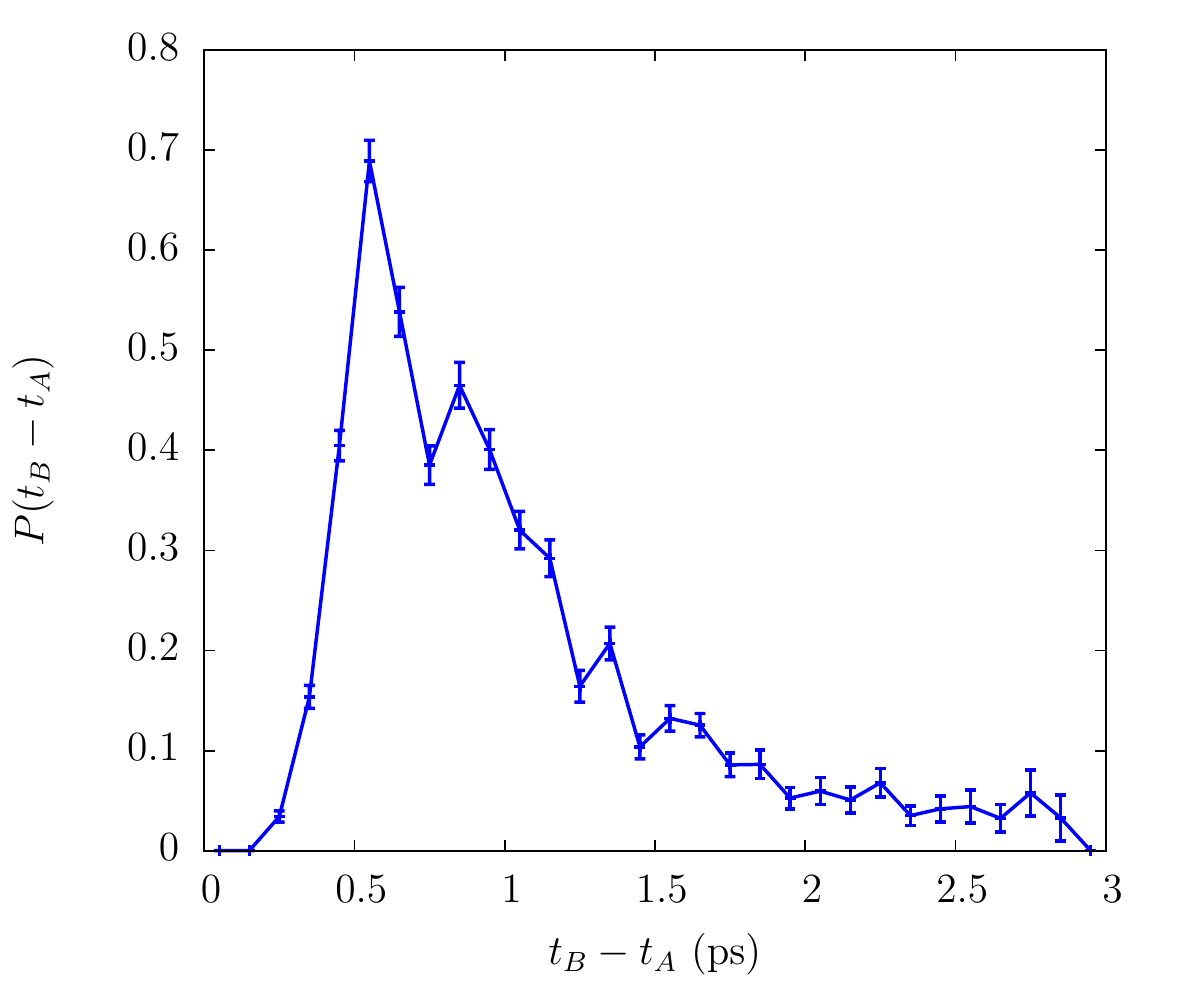}\end{center}
\caption{\label{fig:Evap:tAtoB}Distribution of evaporation event
  durations.}
\end{figure}

\begin{figure}
\begin{center}\includegraphics{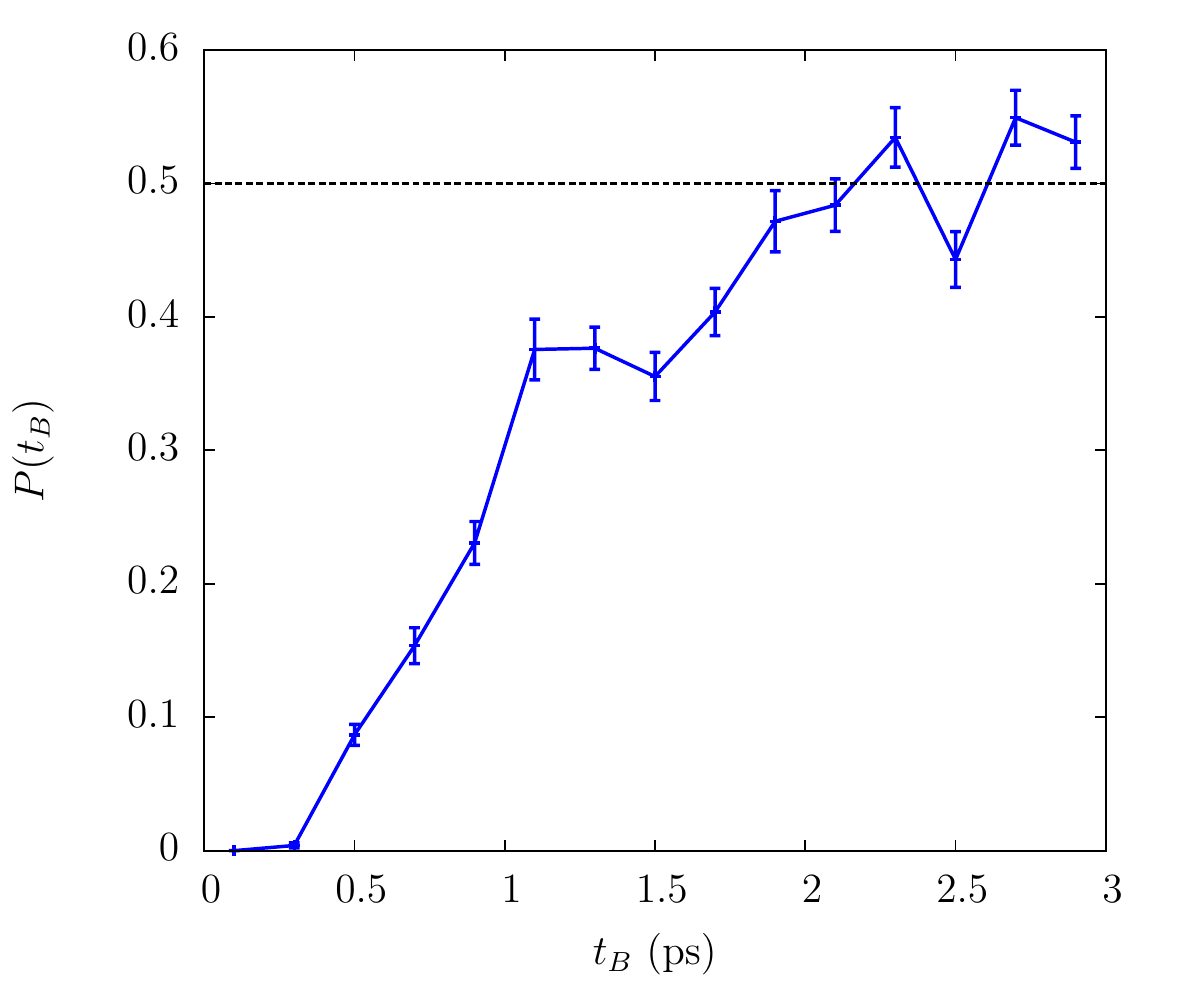}\end{center}
\caption{\label{fig:Evap:tB}Distribution of times at which the
  evaporation event completes.}
\end{figure}

For each trajectory, let $t_A$ be the latest time for which the system is in
basin~$A$, and let $t_B$ be the latest time for which the system is not in
basin~$B$.  These times roughly characterize the points along the trajectory
at which the evaporation event begins and concludes.
Figure~\ref{fig:Evap:tAtoB} shows the distribution of the time difference
$t_B - t_A$.  Most evaporation events take under~$1\,$ps, and very few take
just under $3\,$ps.  Hence, the $3\,$ps trajectory length we chose to use
for our TPS sampling is long enough.  Correcting the distribution of times
$t_B - t_A$ for the bias towards short evaporation events owing to their
larger number of possible starting times does not change this conclusion.
This is demonstrated in Figure~\ref{fig:Evap:tB}, which shows the
distributions of times~$t_B$.  If the TPS trajectory length is sufficiently
long, then this distribution should rise from zero at small~$t_B$ and
plateau to a constant for $t_B$ much larger than the typical time for an
$A$-to-$B$ transition to occur.  This is indeed observed.  Were the TPS
trajectory length too short, there would be no plateau region.

\bibliography{bibl}

\end{document}